\documentclass[aps,twocolumn,showpacs,superscriptaddress,groupedaddress, nofootinbib]{revtex4-1}

\usepackage{hyperref}
\usepackage{multirow}
\usepackage{mathrsfs,amsmath} 
\usepackage{verbatim}
\usepackage{amsmath}
\usepackage{amsfonts}
\usepackage{amssymb}
\usepackage{amsthm}
\usepackage{bm}
\usepackage{xparse}
\usepackage{xcolor}
\usepackage{textcase}
\usepackage{url}
\usepackage{lipsum}
\usepackage{appendix}
\usepackage{dsfont}
\usepackage{epstopdf}
\usepackage{footnote}
\usepackage{tgbonum}
\usepackage{blindtext}
\usepackage{enumitem}
\usepackage{mathrsfs,amsmath} 
\usepackage{float}
\usepackage{subfloat}
\usepackage{wrapfig}
\usepackage{pslatex}
\usepackage{amsmath}
\usepackage{amsfonts}
\usepackage{setspace}
\usepackage{epstopdf}
\usepackage{wrapfig}
\usepackage{csquotes}
\usepackage{hyperref}
\usepackage{graphicx}
\usepackage{amssymb}
\usepackage{multirow}
\usepackage{xcolor}
\usepackage{framed}
\usepackage{mathtools}
\usepackage{tcolorbox}

\definecolor{shadecolor}{rgb}{0.8,0.9,1}
\newcommand{\ket}[1]{| {#1} \rangle} 
\DeclareDocumentCommand{\Tr}{m m O{\big}}{{\rm Tr}_{\:\!{#1}}#3({#2}#3)}

\newcommand{\Q}{\mathbb{Q}}
\newcommand{\N}{\mathbb{N}}

\begin{document}
\title{Potentiality realism: A realistic and indeterministic physics based on propensities}
\author{Flavio Del Santo}
\affiliation{Group of Applied Physics, University of Geneva, 1211 Geneva 4, Switzerland; and  Constructor University, Geneva, Switzerland }
\author{Nicolas Gisin}
\affiliation{Group of Applied Physics, University of Geneva, 1211 Geneva 4, Switzerland; and  Constructor University, Geneva, Switzerland }

\date{\today}

\begin{abstract}
\noindent 
We propose an interpretation of physics named \emph{potentiality realism}. This view, which can be applied to classical as well as to quantum physics, regards \emph{potentialities} (i.e. intrinsic, objective propensities for individual events to obtain) as elements of reality, thereby complementing the \emph{actual properties} taken by physical variables. This allows one to naturally reconcile realism and fundamental indeterminism in any theoretical framework. We discuss our specific interpretation of propensities, that require them to depart from being probabilities at the formal level, though allowing for statistics and the law of large numbers. This view helps reconcile classical and quantum physics by showing that most of the conceptual problems that are customarily taken to be unique issues of the latter -- such as the measurement problem -- are actually in common to all indeterministic physical theories.

\end{abstract}

\maketitle


\section{Introduction}

Centuries of formalization of physics have led to a narrative according to which the universe supposedly evolves through deterministic laws, namely, every event happens as a necessity. This is considered almost a truism in classical physics (Newtonian mechanics, electromagnetism, relativity) and even after the advent of quantum mechanics, many consider the probabilistic prediction of the theory as a mere epistemic concept, while at the fundamental level everything is governed by the deterministic Schr\"odinger equation. Perhaps due to such a strong belief of most of the physicists in determinism, it seems that indeterminism has assumed the status of a bug to be eliminated, and this has led to a series of deep-rooted misconceptions that tend to scramble indeterminism together with a lack of causality or realism, or of lawfulness altogether.

We defend that it is among the essential characteristic of science to provide explanations of natural phenomena, namely the possibility of telling a story about how Nature does it. And this requires to introduce metaphysical elements that causally interact, which provides a strong motivation for (some form of) realism (see Section \ref{onto}). Moreover, we are quantum physicists, and our positions are strongly influenced by the indeterministic worldview brought about by quantum theory, a view rooted in fundamental results such as Heisenberg relations or the violation of Bell inequalities.\footnote{In fact, an experimental violation of Bells inequalities guarantees that, under the assumption of no-signaling (parameter independence), if the measurement settings have an arbitrarily small amount of independence, then the outcomes would be genuinely random \cite{putz} (see also Ref. \cite{delsantogisin1} for a discussion).} We have also gone beyond quantum theory, noticing that believing in determinism even at the classical level is too strong of an assumption, which is not supported by observation and requires to assume infinite information at every point of space(-time). In fact, in a series of recent works, we have proposed alternative, fundamentally indeterministic interpretations of (classical and relativistic) physics \cite{gisin1, delsantogisin1, NGHiddenReals, delsantogisin2, delsanto2021, openpast}.

The aim of this paper is therefore to spell out in some detail what a realistic description of an indeterministic world entails. We will show that there is a natural way of maintaining both realism and indeterminism if one assumes that physical systems are characterized at the fundamental level not only by their \emph{actual possessed properties}, but also by their intrinsic (quantifiable) \emph{potentialities}.\footnote{Note that we take indeterminism to be a sufficient condition, i.e., that there exist at least some events that are fundamentally not predetermined with certainty, about which scientist can accumulate statistics.} Our view drastically differs from traditional realistic positions which uphold objects to be existing and identified by their intrinsic actual properties (at any given time). We maintain instead that physical objects exist and are still characterized by their intrinsic properties which are, however, in general only potential (at any given time), i.e. intrinsic tendencies to actualize. Our view can therefore be called \emph{potentiality realism}.\footnote{This view is similar to some interpretations of the quantum state in terms of potentiality, reminiscent of the Aristotelian \emph{potentia}, hinted at by W. Heisenberg and A. N. Whitehead,  (a discussion on Heisenberg and Whitehead views on potentialities can be found in, e.g., Ref. \cite{potentia}). In more recent years, also N. Maxwell \cite{maxwell},  M. Dorato \cite{dorato}, and M. Suarez \cite{suarez, suarez2} have advocated an irreducible dispositional ontology for quantum mechanics.}

\section{Propensities in an indeterministic world}
\label{prop}

A main feature of an indeterministic world is to be able to produce non-necessary events, i.e., instances where new information that was not previously existing comes into being. This new information could be merely random, devoid of any structure \cite{intuition2}. Underdetermined events, however, may have in general well-determined potentialities of various strengths to realize one out of all the possible events  (examples of events are measurements results like spin up or spin down of a spin-1/2 quantum particle, different faces of a dice, bits 0 or 1 of a random number generator, or the position of a system in one region of space or in its complement, etc.).\footnote{Note however, that events are not necessarily the results of measurements. Some events may happen at a microscopic level that humans may not directly perceive.} As time passes, there are events (either induced by, e.g,  a measurement, or by a spontaneous process) that lead to the realization of one of the potential outcomes, thus creating new information that was not existing before.\footnote{This may require a concept of time different from the parametric time appearing in physical equations, or the geometric time of relativity. See Ref. \cite{delsantogisin2} for a discussion on the compatibility between relativity and indeterminism, and Ref. \cite{timereally} for the two different kinds of parametric and creative time. A similar distinction is also made in Ref. \cite{piron} between a ``historical time'' and a ``parametric time''. These events reduce the indeterminacy of the physical system.}
Hence, such a description requires objective, causally determined -- and yet not ruled by necessity -- intrinsic tendencies for individual events to obtain, i.e., propensities.\footnote{Propensities were originally introduced by K. R. Popper \cite{popper} as an interpretation of probability calculus and then developed into many different variants by a number of authors. For comprehensive reviews of propensity theories see \cite{propensities1, propensities2} and references therein. The view of intrinsic properties with a causal value is shared also by S. Shoemaker who asserts ``that the identity of a property is completely determined by its potential for contributing to the causal powers of the things that have it.'' \cite{schoemaker}.} We deem propensities to be as necessary in a description of a causal  indeterministic world as actual possessed properties are necessary to describe a deterministic world. Note that the physical properties of a system have in general a fundamental (i.e. ontic) indeterminacy \cite{mariani}, and there are events that reduces the indeterminacy (e.g., measurements). This fundamental indeterminacy then affects the future evolution of the system in time, leading to possible alternative future scenarios, i.e., to indeterminism. Therefore, indeterminism is a consequence of intrinsic indeterminacy (even if the dynamical equations may be deterministic). The intrinsic indeterminacy is in turn quantified by a propensity.

In the quite extensive literature devoted to propensities, there have been several proposals to conceive these objective, intrinsic tendencies. In what follows, we will expound our interpretation of propensities, arguing that this is the most suitable one to describe indeterminism in physics while maintaining realism and causality. 

To begin with, our view departs from a major category of propensity theories pursued by several authors, such as I. Hacking \cite{hacking} or D. Gillies \cite{propensities2}, who advocate that propensities are ``long-run'', i.e., they are dispositional properties of long, possibly infinite, series. If this were the case, however, it would entail a sort of ``nonlocality in time'', because if a propensity is to be an element of reality, then it would have to exist as a single entity that influences series of events happening over a span of time that could require the entire age of the universe and its (possibility infinitely extended in time) future. Hence, we reject this view of a time-nonlocal infinite sequence of runs which determines the behavior of individual events, but rather the long-run observed frequencies are manifestation of individual-event propensities. Long-run sequences are nothing but the accumulation of individual events, while the opposite does not follow.
In fact, propensities should characterize objective tendencies of events that can in principle happen only once and not be repeatable (in which case however they are not even approximately measurable). Note that in the case of a single observed outcome, the only information that can be extrapolated about the associated propensity is that it was different from zero.

On the contrary, we define a propensity to be an intrinsic property of objects -- which in physical theories are called \emph{systems} -- to realize a particular (idealized) measurement outcome, i.e., a quantified tendency to realize a particular outcome when subjected to \emph{specific circumstances}. The specific circumstances are the complete set of causally relevant conditions. Our interpretation of propensities is therefore ``single-case'', and represents a form of indeterministic causation aimed at quantifying a degree of objective possibility (which makes propensities \emph{almost} an interpretation of probability but with due conceptual and formal distinctions; see Section \ref{propvsprob}). On this account,  our view is similar to the early ones of J. H. Fetzer \cite{fetzer} and of L. Ballentine \cite{ballentine}, the latter of whom  maintains that a ``propensity refers to a degree of causality that is weaker than determinism.'' \cite{ballentine}. Propensities are therefore to be thought of as ``indeterministic laws'' or ``potential forces'', really existing in the world and causally connecting events but in a looser sense than necessity (which would be the case in a deterministic world). 

More formally, a propensity is described by a mathematical entity (called itself a propensity) that -- given two causally related events (e.g., the hypothetical outcome of a measurement with its  relevant preparation) -- quantifies the strength of the tendency for the effect to happen, given its causally relevant conditions. In this sense, even if the world is characterized by indeterminacy at the fundamental level, events are related by causality.  Indeterminism does not come out of the blue, is not acausal but is the consequence of fundamental indeterminacy.\footnote{Note that the possibility of having fundamental indeterminacy (i.e., at the ontological level) was for long time mostly dismissed in the philosophical literature. However, in recent years, works like \cite{
barnes, millerwor} have formalized this possibility, thereby given new momentum to the acceptance of fundamental indeterminacy.}

In Section \ref{propvsprob}, however, we will see that retaining this causal interpretation of propensities would require them to slightly modify their formalism with respect to mathematical probabilities. And yet, while  they may not be probabilities at the formal level, they share most of the mathematical structure of probabilities, therefore one can still adopt the writing: $p(a| causal \  conditions)$. This is to be interpreted as the intrinsic tendency of the event  $a$ to happen (for instance, the outcome of a possible measurement performed on a system) given the relevant causal conditions that can influence the realization of that outcome (for instance, the settings of an idealized measurement).\footnote{Our view seems to be similar to the interpretation of conditional probabilities in the analysis of causation that J. Butterfield calls ``conditional with a probabilitistc consequent'' \cite{butterfield2}, as endorsed also by D. Mellor and D. Lewis.} But while $a$ is an event in mathematical terms, i.e., an element of a probability space, the causal conditions are not necessarily so. This is clear in quantum mechanics, where the causal conditions to compute the outcome $a$ of a measurement (i.e., an eigenvalue of the observable $A$, corresponding to the eigenstate $\ket a$) are represented by the quantum state $\psi$  -- which is a ray in a Hilbert space and not an element of the probability space -- and the choice of measurement $A$; the probabilities are then computed according to the Born rule as $p(a|\psi, A)=|\langle a|\psi\rangle|^2$. These ``probabilities'' arising in quantum mechanics are, in fact, sometimes called ``generalized probabilities'' (see e.g. \cite{huges}, chapter 8.1) because they are not defined on standard sets of events, as in Kolmogov's axiomatization, due to the incompatibility of canonically conjugate variables (i.e. the algebraic structure of the set of ``quantum events'' is non-Boolean). Hence, ``the functions assigning probabilities to quantum events are, paradoxically, not probability functions at all, at least, not in a Kolmogorov's sense.'' \cite{huges}.  In the next section, we will discuss in detail the difference between propensities and probabilities and their analogies with quantum ``generalized probabilities''.

Furthermore, contrarily to the standard view of (at least classical) physics, the outcomes are not a manifestation of \emph{actual} pre-existing properties that get unveiled. In that standard view, classical states (as well as realistically interpreted quantum states) are collections of actual values (mathematically associated to n-tuples of real numbers) that are assumed to exist well-determined at any instant in time, no matter how far in the past or in the future. In our view, on the other hand, outcomes are dynamically realized -- i.e., they come into existence, thus creating new information -- from an array of mutually exclusive potential outcomes. Obviously, well-determinate outcomes do exist (and every measurement yields one and only one outcome), but only as a particular case when the value of the associated propensity is 1, i.e., when no alternative is possible and the outcome is deterministic (see Section \ref{propvsprob}).  As such, at any instant in time before the event that corresponds to the realization of one (and only one) outcome out of all the possible ones, the elements of reality -- i.e. the state -- is the collection of propensities associated to each possible and in general still indeterminate outcome. Our proposal is therefore a form of realism that can be called, as said, \emph{potentiality realism} as opposed to the standard form of \emph{actuality realism}. The element of reality is now conceived as a collection of potentialities, each of them quantified by an objectively existing propensity.\footnote{A similar position was previously elaborated in Ref.\cite{gisinreal2}.} Let us emphasisze that propensities are possessed by the system in the same way as actual properties are.

Our definition of states in terms of propensities requires some clarification. J. Butterfield defines a physical state ``as a system's maximal
(or `complete') set of intrinsic (or `possessed') properties'' \cite{butt}. Our definition of a state complies with this. Complete means that a state encapsulates the maximal amount of information existing at present about each relevant degree of freedom of the system. But the maximal information is here about potential properties, i.e.  the collection of all the propensities. Note that thinking in terms of propensities does not add to the ``ontological cost'' of a theory (see also Section \ref{onto}): propensities are indeed not directly epistemically accessible, but neither are the postulated actual values of physical properties in standard views of physics. In both cases, the knowledge is obtained through the same operational procedure of repeated measurements and collected statistic that supposedly approximate the underlying element of reality, be it an actual property or a potential one.

Let us turn now to the second desideratum of a physical state, namely that of being a set of intrinsic properties. Intrinsic means in this context that the propensities are properties of the system alone and are not characteristics of experimental arrangements \cite{popper}, nor of chance set-ups \cite{hacking}. D. H. Mellor states that attributing propensities to a chance set-up ``is to remove completely the point of ascribing a disposition as something that is present whether or not it is being displayed. It is to confound propensity with the chance distribution that displays it’’ \cite{mellor}. In fact, intrinsic tendencies (i.e., possessed propensities) can only be approximately revealed by ideal measurements, \emph{in the same way as possessed properties are} (like the value of a physical quantity, e.g., the energy of an atom). Ideal measurements are noise-free measurements that can be immediately reproduced. In the standard view of classical physics, they are characterized by yielding the same result without any variance (because they merely reveal the magnitude of an actual property). Now, let us assume that it is always possible to find a state that makes the propensity for a certain measurement outcome on a system  to take the value 1  (i.e., they yield the result deterministically) and that an ideal measurement of a propensity is independent of the value of the propensity itself. Then one can use the state corresponding to a deterministic outcome (propensity 1) to characterize the ideal measurement also for any other state. 

Note that, additionally, the collection of all the propensities -- i.e., a state in this ``potentiality'' interpretation -- may have its own well-defined mathematical structure, hence it might be possible to summarize this collection of propensities in a compact mathematical form.

Finally, it ought to be recalled that the violation of Bell inequalities -- by now a well-corroborated empirical fact -- forces us to reject locality (in space). This obviously affects  propensities as well, which have to be in general nonlocal elements of reality, i.e. they cannot be separable at the formal level. Therefore, while we reject the view that propensities have an arbitrary extension in time (i.e., they enjoy locality in time), the violation of Bell inequalities -- while upholding single outcomes and free choice -- forces us to have propensities nonlocal in space. Admittedly, this is in tension with the relativistic worldview, although we have already hinted at ways to render these views compatible with each other in Ref. \cite{delsantogisin2}.  Namely, a single objective tendency quantifies the bias towards the possible realization of correlations between distant indeterministic outcomes (a more detailed discussion can be found in Ref \cite{delsantogisin2}, also in relation to special relativity).

\section{Propensities vs. probabilities: \\Humphreys' paradox }
\label{propvsprob}

In the previous section, we have defined propensities as being intrinsic, objective tendencies that causally \emph{and} indeterministically quantify the possibility of a system to realize an event, given its causally relevant conditions. 
 This seems \emph{prima facie} exactly what a conditional probability describes, if one is to interpret uncertainty not as epistemic but as a manifestation of fundamental indeterminacy. However, a result known as ``Humphreys' paradox'' \cite{humphreys} has posed serious limitations to directly interpreting propensities as objective probabilities. Ballentine rightly calls ``Humphreys’s result a \emph{theorem}, rather than a \emph{paradox} because it is a result that validly follows from its assumptions.'' \cite{ballentine}. In fact, we contend that this can be casted in the form of a no-go theorem which states that the conjunction of the following assumptions is untenable (i.e., leads to contradiction): 
\begin{enumerate}
\item[(i)] Propensities quantify indeterministic causal connections.
\item[(ii)] Propensities are probabilities, i.e., they are defined through all Kolmogorov's axioms (in particular those from which Bayes' rule is derived).
\end{enumerate}

To see this, assume that propensities are indeed probabilities -- assumption (ii). Consider two causally related events, namely, if one of them -- the cause, $C$ -- obtains, it influences the tendency of another event -- the effect, $E$ -- to obtain. This means, by assumption (i), that there is a propensity $P(E|C)$ that quantifies the tendency of $E$ to happen given that $C$ has occurred. Let this propensity connect the event to the cause non-trivially, i.e., $P(E|C)\neq P(E)$. At the same time, since we have assumed propensities to encapsulate the concept of causal connections, whether the effect $E$ will happen or otherwise (i.e., the complementary event $\overline{E}$ happens), it cannot influence the cause, i.e., $P(C|E)=P(C|\overline{E})=P(C)$.\footnote{Causality and time are two intimately related concepts, for causes are happening \textit{before} their effects. Whether causality is fundamental and the concept of time is derived from it, as suggested by H. Reichenbach \cite{reichenbach}, or it is time to be fundamental, as supported by J. Lequyer (see Ref. \cite{timereally}), will not be determinant for the argument and will not be discussed further in what follows.} Note that this is the standard definition of causality -- also knows as no-signaling from the future -- in operational formulations of physical theories (such as in ``generalized probability theories'', GPTs \cite{gpt}), too, where the choice of a measurement cannot influence the state preparation (see Ref. \cite{dariano}).  By assumption (ii), Bayes' rule ought to hold. This means that one can write the conditional probability in terms of its reversed one as (Bayes' rule):
\begin{equation}
P(E|C)=P(C|E)\frac{P(E)}{P(C)}.
\end{equation}
However, plugging in the causality constraint above yields to $P(E|C)=P(E)$, which is in direct contradiction with the assumption of non-triviality of the cause. 

Dropping assumption (i) would lead to accept that propensities are not single-case tendencies but long-run (as defended by Hacking \cite{hacking} and Gillies \cite{propensities2}), but we have already argued against this stance in the previous section. On the other hand, if one wants to maintain assumption (i), propensities need to escape Humphreys' theorem by departing from formal probabilities. Axiomatizations of propensity calculus -- not reducible to Kolmogorov's axioms of probability theory (or equivalent) -- have been proposed by  Fetzer and Nute \cite{nute, fetzer}, one of the present authros (NG) \cite{gisinprop1, gisinprop2}, and Ballentine \cite{ballentine}. 

Note that if propensities are to be compatible with statistical observations, their axiomatization cannot be fully arbitrary and consequently they cannot be arbitrarily different from probabilities, for, in turn, the latter are also constructed to be the limiting case of observed frequencies. That is why  Gillies proposes ``to speak of the probabilistic causal calculus [i.e., axiomatized propensity calculus] as a non-Kolmogorovian probability theory by analogy with non-Euclidean geometry.'' \cite{propensities2}. In fact, in the same fashion that non-Euclidean geometry maintains  most of the structure of standard geometry by only rejecting Euclid's fifth postulate, propensity calculus (non-Kolmogorovian probability) drops the Kolmogorov's axiom(s) that lead to the derivation of the Bayes' rule.
In particular, as already remarked in Ref. \cite{ballentine}, a desideratum for propensities (which is also true for probabilities) is that they obey \emph{Bernoulli's Law of Large Numbers}:
\begin{equation}
\label{lln}
P\left(\left|f_n - p\right|\geq \epsilon \right)\rightarrow 0,
\end{equation}
for every $\epsilon>0$ and for the limit of $n \rightarrow \infty$. In words, $P$ is the probability that quantifies the correlation between the relative frequency of the occurrence of the considered outcome, $f_n$, and the propensity $p$ for that event, which gets stronger as the number of trials $n$ increases.\footnote{Note that Bernoulli's Law relies on some concept of independence of events (see also Ballentine \cite{ballentine}).} This result is derived in standard probability theory, but there $p$ is taken to be a probability as well instead of a propensity. Bernoulli's law of large numbers is therefore at least a necessary element in common between probabilities and propensities, for both of them should be the limit of relative frequencies in long series of measurements. However, at the conceptual level the difference is tremendous: while probabilities only encapsulate \emph{correlations} (which do not imply causation), propensities \emph{cause} relative frequencies, which, in turn, are the observable manifestation of propensities and therefore the operational way to epistemically access them (to an arbitrary approximation). This is why Bayes' rule is valid for probabilities but not for propensities. 

Another desideratum for propensities -- also in common with probabilities -- is that they ought to be bounded between 0 and 1, because they have to account for impossibility and certainty, respectively. But again, due to their causal nature, a propensity of 0 or 1 does not only mean certainty -- of failure or occurrence of the considered event, respectively -- in a measure theoretic sense (as it is in probability theory, where events with probability 0 are not logically excluded but their occurrences form a subset of measure zero). Rather, a propensity of 1 means a necessary causal connection, i.e. \emph{deterministic causation} between the conditions and the occurrence of an event. On the other hand, an intermediate value of a propensity, strictly larger than 0 and smaller than 1, should be instead interpreted as a non-deterministic ``potential  force''. A propensity with value $1/k$ represents a truly unbiased random event for the property of a system that can display $k$ mutually exclusive outcomes. 

Since, however, measurements always lead to a single observed outcome, this necessarily begs the question of \emph{how} the potentialities become actual, i.e. what ``mechanism'' makes propensities evolve from an intermediate value, between 0 and 1, to either 0 or 1 at the time of measurements (and not necessarily only at measurements). This is exactly the analogous of the notorious ``quantum measurement problem'' (in Refs. \cite{delsantogisin1} and \cite{delsanto2021} we discussed this calling it the ``classical measurement problem''), which is a general characteristic of all fundamentally indeterministic theories. At the formal level, in quantum mechanics the measurement problem can be cast as an incompatibility between the linear, unitary, deterministic evolution of the quantum state (Schr\"odinger equation) and the observation of a single outcome upon measurement with the ``collapse'' of the quantum state on one of the eigenstates of the operator corresponding to the measurement. Note that in an indeterministic interpretation of classical physics this can also be regarded in the same terms: the indeterminate state (i.e., the collection of propensities) is bounded by a finite region of phase space which gets deterministically mapped, through the equations of motion (Newtons laws), to an arbitrarily large region of phase space (due to chaotic systems). When a measurement is performed, however, a localization is expected in the same fashion as a ``collapse'' in quantum physics. Hence, in both (indeterministic) classical mechanics and quantum mechanics this actualization of the potentialities requires to look for a ``mechanism'' that ``forces'' the propensities to get determined at measurements. 

The two possible mechanisms that can lead to this is that propensities either evolve into determined outcomes spontaneously -- which is reminiscent of objective collapse models in quantum theory \cite{grw, gisincollapse} -- or if a measurement somehow ``imposes'' a determination -- similarly to the Copenhagen interpretation of quantum mechanics in which an observer or a measurement apparatus (not described within quantum theory) makes the wave function ``collapse''. The former view seems the most compatible with reductionism, whereas the latter one seems to imply a form of a top-down causation \cite{topdown}, where a higher level of description ``forces'' the actualization of the lower level.  Furthermore, the fact that only one outcome can be realized in a given experiment is itself a (metaphysical) assumption. In fact, in principle one can think of a radical interpretation of propensities, in which the possibilities are forever possible (i.e., in our parlance, there is no event that actualizes only one of them, ruling out the others) and the state is the collection of the all the propensities for any possible event of the universe and extended everywhere in time. This is the analogous of the ``many-worlds interpretation'' of the quantum state and it shares with it its enormous ontological baggage (see Section \ref{onto}). Finally, we notice that the standard deterministic interpretation of classical physics is analogous to adding unobservable ``hidden variables'' (i.e., the real numbers as the actual value of physical variables, see \cite{NGHiddenReals}), in the same way that Bohmian mechanics adds hidden variables to the quantum state (i.e. the position of each quantum particle) to deterministically complete the theory.

Upon repetitions, under the same causally relevant conditions, one can measure with arbitrary precision the  propensity of a system, and approach it in the limit. If the value of the propensity is 1, the system will always display the same particular outcome -- under the assumption of ideal measurements (see Section \ref{prop}). Note that -- according to standard classical physics -- to measure the supposedly existing actual value of some physical quantity of a system  or to measure the (also supposedly existing) propensity for a system to display some particular result -- as postulated in potentiality realism -- one has to accumulate enough statistics. Indeed, when we assert that a system possesses (at present) a certain property, e.g., that some physical quantity has some property $a$, we mean that if we would test for this property, i.e. we perform a (ideal) measurement of that physical quantity,  we would find with certainty this property $a$. In practice, however, such tests need to be reproduced many times until noise and false positive can be dealt with. This is very similar to the case when we assert that a system possesses (at present) a certain propensity.

It is in this statistical analysis of measurements that standard Kolmogorov probabilities play their role. This is very well understood by the mathematical measure theory (in terms of probability spaces, Borel sets, etc). However, and this is crucial, there is a priori no reason to believe that sets of statistics corresponding to different measurements can be combined into a single probability space. This is not just a theoretical hypothesis, but it is actually demonstrated by quantum theory. Hence, propensities of a system to display results corresponding to different physical quantities, i.e., different measurement settings, should not be expected a priori to satisfy Kolmogorov's axioms -- in compliance with Humphreys' theorem. In fact, if there were a global probability space for all propensities $p(a|measurement \ conditions)$, then the joint probability $p(a, measurement \ conditions)$ would be well-defined and this would imply the existence of ``hidden variables'' -- the elements of the global probability space -- that provide a deterministic description. Then one is left with two possibilities:
\begin{enumerate}
\item If the hidden variables are accessible, then this world is not really indeterministic, or
\item If the hidden variables are fundamentally inaccessible, then this world would require to be described through variables that are intrinsically non-physical (like, e.g., real number in classical physics or exact positions in Bohmian mechanics).
\end{enumerate}

A clarification is here in order. In the second statement, fundamentally inaccessible has a stronger meaning than the inaccessibility of actual values of standard classical physics or of the propensities in our proposal of potentiality realism. The latter two, in fact, while effectively inaccessible, can be approached by successive approximations in repeated measurements (by means of Bernoulli's law of large numbers). The hidden variables of the global probability space instead would be as inaccessible as actual positions in Bohmian mechanics, and have therefore a much less justifiable physical ground.\footnote{This treatment in terms of hidden variable has some resemblance with the solution to the Humphreys' theorem proposed by D. Miller \cite{miller} and the late Popper \cite{popper90}. They maintain that propensities can only be attributed to the whole causal past (the past light cone in relativistic terms or the entire state of the universe) of an event: ``The non-standard conditionalization is on an ‘event’ outside the probability space—the entire present state of the universe—which determines the probabilities of the events in the probability space'' \cite{propensities1}. We notice that Bell nonlocality jeopardizes the view that all the relevant causal connection  lie only in the past light cone of a local event. Instead, one should take into account the compound of the past light ones of all the systems entangled with the one under consideration.}

%

\section{Finite information quantities}
\label{FIQs}

In Ref. \cite{delsantogisin1}, we have proposed an alternative, indeterministic interpretation of classical physics, based on propensities. Starting by noticing that physical variables are usually assumed to have an actual value, determined with infinite precision (encapsulated by a real number), we assumed instead as a fundamental principle the \textit{finiteness of information density}, namely that finite volumes of space(-time) can contain only a finite amount of information (see also \cite{gisin1}). We have thus introduced a model of classical physics in which physical variables are assumed to take values, instead of in the real numbers, in a new class of mathematical entities that we named \textit{finite-information quantities} (FIQs). 
To illustrate this, let us start from the standard view. Let  $\Gamma$ be a physical quantity (say the position of a particle) that take values in the unit real interval, i.e., $\Gamma \in [0,1]$ and let us write it in binary base:
\begin{equation*}
\Gamma=0.\gamma_1\gamma_2\cdots \gamma_j \cdots,
\end{equation*}
where the bits $\gamma_j\in\{0,1\}$, $\forall j\in \N^+$. Because $\Gamma$ is a real number, its infinite bits are all given at once with a well defined value 0 or 1. In our model, on the other hand, we impose that the information should be finite, such that not all the digits should be determined at all times. However, since we require it to be empirically equivalent to standard classical physics, the first (more significant and perhaps known) $n$ bits should be fully determined at time $t$:
\begin{equation*}
\Gamma \left(n(t)\right)=0.\gamma_1\gamma_2\cdots \gamma_{n(t)} ?_{n(t)+1}\cdots ?_k\cdots,
\end{equation*}
where each bit $\gamma_j\in\{0,1\}$, $\forall j\leq n(t)$, and the symbol $?_k$ here means that the $k$th digit is a not yet actualized binary digit, but only its propensity exists before time $t$. For each digit $j$ of a physical quantity $\Gamma(n(t))$, we associate the propensity $q_j\in [0,1] \cap \Q$ that quantifies the tendency of the $j$-th binary digit to take the value 1, such that it is $q_j=1$ iff the $j$-th bit is \textit{certainly} 1 and $q_j=0$ iff the $j$-th bit is \textit{certainly not} 1 (i.e. if it is 0).

A FIQ is then defined as an ordered list of propensities $\{ q_1, q_2, \cdots , q_j, \cdots \}$, that satisfies the (necessary) condition: $\sum_j I_j < \infty$, where $I_j=1-H(q_j)$ is the information content of the propensity, and $H$ is the binary entropy function of its argument. This ensures that the information content of FIQs is bounded from above. A physical quantity $\Gamma$ thus reads:
\begin{equation*}
\Gamma (n(t))=0.\underbrace{\gamma_1\gamma_2\cdots \gamma_{n(t)}}_{q_j\in \{0, 1\}}\overbrace{?_{n(t)+1}\cdots }^{ q_k\in(0, 1)}.
\end{equation*}
Since most of classical systems are chaotic, this fundamental indeterminacy (here modeled with FIQs), inevitably leads to indeterminism in classical physics too (see Ref. \cite{gisin1}). Therefore, FIQs render classical physics indeterministic while upholding a concept of physical state in potentiality realism (see Section \ref{prop}). Indeed, a pure state (i.e. containing the maximal information possible about a system) is here a list of propensities with the constraint that the information therein contained is always finite.

\section{Why Realism? Benefits and costs of an ontology}
\label{onto}

In the previous sections, we have defended the position that propensities -- at least in their causal, single-case interpretation advocated for here -- allow to maintain realism and at the same time fundamental indeterminism. One may, however, ask why one should look for a realistic account of physics in the first place. This question is perhaps one of the most contended in the history of philosophy of science, (for an overview of the main arguments for and against scientific realism see, e.g., Ref. \cite{sep} and references therein ) so it would obviously be impossible, and completely beyond the scope of this short paper, to even try to provide an account of the debate surrounding justifications and criticisms of metaphysical realism. Here we will just quickly remind the reader of why it seems fruitful to consider realistic positions; moreover, we will argue that if one assumes a realistic worldview an ontology based on propensities is the most ``economical''.

A way to justify realism is to regard science, and physics in particular, as a form of knowledge characterized by both \emph{explanatory power} and \emph{predictive power}. The latter is uncontroversially considered a signature feature of science and according to strong empiricists and instrumentalists science should not strive for anything more than such a feature (see, e.e, \cite{vanfrassen}). This view, however, does not seems to bring particular insights on nature and more importantly it precludes the meaningfulness of certain questions (a prime example is that of Bell's inequalities that arguably would not have been derived within a fully instrumental use of quantum theory). One thus wants science to be provided with explanatory power, namely, the ability to tell consistent stories about \emph{how Nature does it}. This requires to charge our theories with metaphysical elements because in order to give explanations one has to postulate the existence of entities in the world that interact and causally account for the observed phenomena. Obviously, this alone would not be science, for also mythology or religion aim to explain natural phenomena, but in a more arbitrary way (for example, a stormy sea could be explained by the god Poseidon, or the thunder bolt by the god Zeus who are in anger). So, in a satisfactory scientific theory,  the metaphysical elements should be built as an interplay between the observations and the predictions of a theory, on the one hand, and the underlying \emph{things of the world} on the other, i.e.,  postulated elements of reality  (these can be fields, particles, specific degrees of freedom with their actual magnitudes, propensities, etc.).\footnote{Physicists may be acquainted with the definition of an ``element of reality'' as defined by Einstain, Podolski and Rosen in their influential EPR paper, in terms of perfect predictability of physical quantities. Our use of the term ``element of reality'' here does not refer to that definition and it simply intuitively refers to the ``things'' that exist in the world.} A necessary condition for these underlying elements of reality is that, while being in general not directly observable, they are compatible with -- i.e. they do not contradict -- the predictions of a theory and actual observations. These things of the world form the ontology of a theory, which provides the foundations of realism at least as a working hypothesis.

Note that there are many (in principle infinite) different possible ontologies compatible with the same theory and set of observations at any given time. One should therefore define standards or guidelines to adopt one ontology among all the possible compatible ones. As examples of possible ontologies consider, for instance, the one theorized by the pre-Socratic philosopher Anaxagoras who maintained that everything is composed of ``seeds'' (\emph{spermata}), namely microscopic and fully formed versions of any observed substance and beings (e.g., miniaturized versions of humans, stones, horses, trees, etc.)  and that ``in everything there is a portion of everything'' \cite{lewis}. More recently, the so-called many-worlds interpretation of quantum theory postulated the existence of a multiplicity of (possibly uncountably infinitely many) worlds, namely that for each and every quantum measurement the whole universe, with its tremendous complexity, splits  into many copies thereof that only differ from each other by the result of the quantum measurement.
What unites these two ontologies is that, although they provide an explanation, they are extremely ``costly'', for they both imply an absurd inflation of required things of the world: Anaxagoras in terms of types of component, many-wolds in terms of amount of universes (so an inconceivable amount of information, etc.) that form reality.

This suggests that one standard to prefer an ontology over its competitors is to favor more ``economical'' ontologies, given the same compatibility with the observations and predictions of a theory.\footnote{This resembles to a certain extent E. Mach's ``economy of science''. However, he went so far as to state that science should be stripped away from any metaphysics and our theory should just arrange data in the most economic way. As discussed above, this does not seem satisfactory to us in so far as alone it does not seem to provide explanations.} Therefore, if the observed phenomena can be explained resorting to fewer types or a smaller  amount of elements of reality, that ontology should be preferred.\footnote{Note that this is not necessarily  an argument in favor of physical reductionism, which, as a matter of fact, we do not support. The aim of science should not be to  unify everything into a single entity that explains everything, and there could be a variety of types of entities, dynamical laws, etc. that are all necessary to explain different phenomena.}

Now, it is generally known that quantum physics is the most successful theory and that its predictions are only probabilistic. A more thoughtful analysis of classical physics also leads to conclude that at the observational level also classical outcomes are only known with an interval of confidence that is characterized by statistical repetitions from which one extrapolates a probability distribution. And chaotic systems make the indeterminacy grow exponentially as time passes. (Note that the standard story of classical physics explains away probabilities by assuming that there exists actual, infinitely precise determinate values that are approximately revealed by measurements, but this is already an ontological stance). Therefore, in the actual practice of science, measures of indeterminacy, i.e. probabilities, play not \emph{a} but \emph{the} major role. Thus the most economical ontology is to assume that probabilities are manifestations of an underlying ontology that resembles them in structure, i.e. indeterministic forces in the form of causal propensities as expounded in this paper.

\section{discussion}
We have discussed an interpretation of physics that we called potentiality realism. While not being tied to any particular physical theory, it helps  clarify the rather generally misunderstood fact that realism and fundamental indeterminism are compatible metaphysical properties. Potentiality realism grants to intrinsic tendencies or propensities, the role of ``elements of reality''. A physical state is thus not characterized by its actual properties (the postulated values taken by each variable corresponding to each relevant degree of freedom), but by potential, non-deterministic ``forces'' that quantify the intrinsic tendency of a property (as opposed to its possible alternative) to actualize. 

We would like to emphasize the picture of indeterminism in physics that our view presents. Here, indeterminism is not represented by stochastic fluctuations in the dynamical evolution, nor by arbitrarily acausal ``jumps''; here no causal chain starts from nowhere. In our view, an indeterministic evolution is the causal consequence of fundamental indeterminacy. As time passes, new information is created and thus the indeterminacy is reduced. Depending on the dynamical system, the evolution is then driven by the fresh information, one way or the other. This view of indeterminism is not acausal, simply because, as time flows, some potentialities get excluded.

While our view allows to regard the world as indeterministic while maintaining a clear causal structure, a number of issues remain unsettled. For instance, how do potentialities become actual? Is it a spontaneous process, or does this pave the way to think in terms of non-reductionism in which different domains of reality need to act on each other to provide an explanation? Furthermore, in this view, states are not collections of actual properties, but of propensities (i.e., potential values, of which the actual ones are a subset, corresponding to a unit propensity). But propensities dynamically change their values, eventually becoming 0 or 1 when the outcome of an experiment obtains. But then, according to what laws do the propensities evolve? Do they require a second-order dynamics?

By posing these challenging problems, potentiality realism provides a general conceptual framework to analyze the problems of quantum theory in relation to its classical counterpart. In fact, it partly deprives quantum physics of its uniqueness with respect to most of its conceptual issues and ``mysteries''. It shows that what is peculiarly attributed to quantum mechanics was in the large part the result of a historical contingency: Fundamental indeterminacy, the measurement problem, 
 a plurality of interpretations (including a ``many-world one'' and Bohmian mechanics) are all present in a potentiality realistic interpretation of \emph{any} physical theory, being it classical or quantum (or any hypothetical post-quantum theory). What remains unparalleled in quantum physics is the existence of incompatible variables, which is  in fact the only place where the Plank constant $h$ appears (and in Bell nonlocality, which requires measurements in incompatible bases to violate the classical local bound \cite{bellinc}).

Interpreting quantum mechanics remains one of the greatest challenges of modern science. But if one thinks twice, this challenge lies to a large extent above and beyond quantum physics and, hidden behind historically rooted dogmatisms, the great challenge has always been to interpret physics \emph{tout court}.

\acknowledgements
This research was supported by the FWF (Austrian Science Fund) through an Erwin Schr\"odinger Fellowship (Project J 4699: Revisiting the foundations of probability theory for the description of an indeterministic physics), and the Swiss National Science Foundation via the NCCR-SwissMap. We thank Jeremy Butterfield, Claudio Calosi, Federico Laudisa, Baptiste Le Bihan, Christian W\"uthrich and David Miller for useful comments.

\begin{small}

\end{small}


\begin{thebibliography}{200}

\bibitem{delsantogisin1} Del Santo, F. and Gisin, N. (2019). Physics without determinism: Alternative interpretations of classical physics. \emph{Physical Review A}, 100(6), 062107.
\bibitem{gisin1} Gisin, N. (2019). Indeterminism in Physics, classical chaos and Bohmian mechanics. Are real numbers really real?. \emph{Erkenntnis}, 86(6), 1469-1481.

\bibitem{NGHiddenReals} Gisin, N. (2019). Real numbers as the hidden variables of classical mechanics. In \emph{Quantum Studies: Mathematics and Foundations}, 7(2), 197-201.
\bibitem{delsantogisin2} Del Santo, F. and Gisin, N. (2021). The relativity of indeterminacy. \emph{Entropy} 23, 1326.
\bibitem{delsanto2021} Del Santo, F. (2021). Indeterminism, causality and information: Has physics ever been deterministic?. In \emph{Undecidability, Uncomputability, and Unpredictability} (63-79). Cham: Springer.
\bibitem{openpast} Del Santo, F. and Gisin, N. (2023). The open past in an indeterministic physics. \emph{Foundations of Physics}, 53(1), 1-11.



\bibitem{putz} P\"utz, G., Rosset, D., Barnea, T.J., Liang, Y.C. and Gisin, N. (2014). Arbitrarily small amount of measurement independence is sufficient to manifest quantum nonlocality. \emph{Physical Review
Letters}, 113(19), 190402

\bibitem{intuition2} Gisin, N. (2021). Indeterminism in physics and intuitionistic mathematics. \emph{Synthese}, 1-27.

\bibitem{potentia} Lestienne, R. (2022). \emph{Alfred North Whitehead, Philosopher of Time}. London: World Scientific. 

\bibitem{maxwell} Maxwell, N. (1988). Quantum propensiton theory: A testable resolution of the wave/particle dilemma. \emph{The British Journal for the Philosophy of Science}, 39(1), pp.1-50.
\bibitem{dorato} Dorato, M. (2006). Properties and dispositions: some metaphysical remarks on quantum ontology. In \emph{AIP Conference Proceedings} (Vol. 844, No. 1, pp. 139-157). American Institute of Physics.
\bibitem{suarez2} Suárez, M. (2004). Quantum Selections, Propensities and the Problem of Measurement. \emph{British Journal for the Philosophy of Science}, 55(2).
\bibitem{suarez} Suárez, M. (2007). Quantum propensities. \emph{Studies in History and Philosophy of Modern Physics}, 38(2), pp.418-438.


\bibitem{timereally} Gisin, N. (2017). Time really passes, science can’t deny that. In \emph{Time in physics} (pp. 1-15). Cham: Birkh\"auser.
\bibitem{piron} Horwitz, L. P., and Piron, C. (1973). Relativistic dynamics. \emph{Helvetica Physica Acta}, 46(3), pp.316-326.


\bibitem{popper} Popper, K. R. (1959). The propensity interpretation of probability. \emph{The British Journal for the Philosophy of Science}, 10(37), 25-42.
\bibitem{propensities1} Berkovitz, J. (2015). The propensity interpretation of probability: A re-evaluation. \emph{Erkenntnis}, 80(3), 629-711.
\bibitem{propensities2} Gillies, D. (2012). \emph{Philosophical theories of probability}. London: Routledge.
\bibitem{schoemaker} Shoemaker, S. (1980). Causality and Properties. In: Van Inwagen, P. (eds) \emph{Time and Cause. Philosophical Studies Series in Philosophy}, vol 19. Dordrecht: Springer.


%

\bibitem{mariani} Calosi, C. and Mariani, C. (2021). Quantum indeterminacy. \emph{Philosophy compass}, 16(4), 12731.

\bibitem{hacking} Hacking, I. (1965). \emph{Logic of statistical inference}. Cambridge: Cambridge University Press.
\bibitem{fetzer} Fetzer, J. H. (1981). \emph{Scientific knowledge: Causality, explanation and corroboration}. Dordrecht: Reidel.
\bibitem{ballentine} Ballentine, L.E. (2016). Propensity, probability, and quantum theory. \emph{Foundations of physics}, 46(8), 973-1005.


\bibitem{barnes} Barnes, E. and Williams, J. R. (2011). A Theory of Metaphysical Indeterminacy. In Karen Bennett, K, and Zimmerman, D. (eds.), \textit{Oxford Studies in Metaphysics} Vol. 6. Oxford: Oxford University Press.
\bibitem{millerwor} Miller, M.E. (2021). Worldly imprecision. \emph{Philosophical Studies}, 178(9), pp.2895-2911.
\bibitem{butterfield2} Butterfield, J. (1992). Probabilities and conditionals: Distinctions by example. In \emph{Proceedings of the Aristotelian Society} (Vol. 92, pp. 251-272).  Ney York: Wiley.

\bibitem{huges} Hughes, R.I. (1989). The structure and interpretation of quantum mechanics. Cambridge: Harvard University Press.

\bibitem{gisinreal2} Gisin, N. (2015) A Possible Definition of a Realistic Physics Theory. \emph{Internation Journal of Quantum Foundations}, 1(1), 18-24.

\bibitem{butt} Butterfield, J. (2005). Determinism and indeterminism. \emph{Routledge Enciclopedia of Philosophy}. London: Routledge.


\bibitem{mellor} Mellor, D. H. (1971). \emph{The matter of chance}. Cambridge: Cambridge University Press.

\bibitem{humphreys} Humphreys, P. (1985). Why propensities cannot be probabilities. \emph{The philosophical review}, 94(4), 557-570.

\bibitem{reichenbach} Reichenbach, H. (1956). \emph{The direction of time}. Berkeley: University of California Press.




\bibitem{gpt} Plávala, M. (2021). General probabilistic theories: An introduction. \emph{arXiv preprint}:2103.07469.

\bibitem{dariano} Chiribella, G.,D’Ariano, G.M. and Perinotti, P. (2016). Quantum from principles. In \emph{Quantum theory: informational foundations and foils}. Dordrecht: Springer.
















\bibitem{nute} Fetzer, J.H. and Nute, D.E. (1980). A probabilistic causal calculus: Conflicting conceptions. \emph{Synthese}, 241-246.

\bibitem{gisinprop1} Gisin, N. (1984). Propensities and the state‐property structure of classical and quantum systems. \emph{Journal of mathematical physics}, 25(7), 2260-2265.
\bibitem{gisinprop2} Gisin, N. (1991). Propensities in a non-deterministic physics. \emph{Synthese}, 89(2), 287-297.



\bibitem{grw} Ghirardi, G. C., Rimini, A., Weber, T. (1986). Unified dynamics for microscopic and macroscopic systems. \emph{Physical Review D}, 34(2), 470.
\bibitem{gisincollapse} Gisin, N. (1989). Stochastic Quantum Dynamics and Relativity. \emph{Helvetica Physica Acta}, 62(4), 363–371

\bibitem{topdown} Drossel, B. and Ellis, G. (2018). Contextual wavefunction collapse: An integrated theory of quantum measurement. \emph{New Journal of Physics}, 20(11), p.113025.





\bibitem{miller} Miller, D. 1991. Single-case probabilities. \emph{Foundations of Physics}, 21(12), 1501–1516.

\bibitem{popper90} Popper, K. R. (1990). \emph{A world of propensities}. Bristol: Thoemmes.



\bibitem{sep} Chakravartty, Anjan, "Scientific Realism", The Stanford Encyclopedia of Philosophy (Summer 2017 Edition), Edward N. Zalta (ed.), URL = <https://plato.stanford.edu/archives/sum2017/entries/scientific-realism/>.
\bibitem{vanfrassen} van Fraassen, Bas C. (1980). \emph{The Scientific Image}, Oxford: Oxford University Press.

\bibitem{lewis} Lewis, E. (2000). Anaxagoras and the Seeds of a Physical Theory. \emph{Apeiron}, 33(1), 1-24.

\bibitem{bellinc} Wolf, M. M., Perez-Garcia, D., and Fernandez C. (2009). Measurements incompatible in quantum theory cannot be measured jointly in any other no-signaling theory. \emph{Phys. Rev. Lett.} 103, 230402.























\end{thebibliography}
\end{document}